\documentclass[12pt]{article}

\usepackage{epsfig}
\usepackage[T2A,OT1]{fontenc}
\usepackage[koi8-r]{inputenc}
\usepackage{amssymb}

\def\be{\begin{equation}}  
\def\ee{\end{equation}}   
\def\bea{\begin{eqnarray}} 
\def\eea{\end{eqnarray}} 
\def\l{\label}

\def\nn{\nonumber \\}

\def\qq{\qquad}
\def\cqq{\;, \qquad}
\def\dqq{\;. \qquad}

\def\o{\over}
\def\le{\leqslant}

\def\ba{\begin{array}}
\def\ea{\end{array}}

\renewcommand{\d}{\mathrm{d}}

\title{Total cross section of  
$ p p \rightarrow K^+\Lambda  p$ \\ reaction near threshold}  

\author{O.Grebenyuk\footnote{e-mail: oleg@mail.cern.ch} \\ \\
Petersburg Nuclear Physics Institute}
\date{}

\begin{document}

\large

\maketitle

\large

\begin{abstract}

 The total cross sections of the $ p p \rightarrow K^+\Lambda  p$
reaction near threshold measured at the COSY(Juelich) synchrotron
using the internal target facility COSY-11 are described in the
frame of the model taking into account the one pion and one kaon
exchange as well as the final state interaction
mechanisms. It is shown that near the threshold all these
mechanisms give the compatible contributions, but with the 
increasing energy the one kaon exchange mechanism becomes
the dominating one.

\end{abstract}

\normalsize
\section*{Introduction}

The total cross sections of the $pp \to K^+\Lambda  p$
reaction near threshold measured at the COSY(Juelich) synchrotron
using the internal target facility COSY-11 
\cite{Balewski1998,Sewerin1999,Kowina2004} 
present good test for the various models of the associated strangeness 
production. In Tab. \ref{tab:tb_vs_q_pp} the 13 values of 
$\sigma_{tot} (pp \to K \Lambda p)$ measured at different periods 
\cite{Balewski1998,Kowina2004} are presented.
\begin{table}[ht]
\centering
\begin{tabular}{|c|c|c|c|} 
   \hline 
   $Q_{pp}$, GeV      &  $T_p$, GeV  &  
$\sigma_{tot} (pp \to K \Lambda p)$ nb  & Reference \\  \hline
0.00068  &   1.58377   & 2.1 $\pm$ 0.2  & \cite{Balewski1998}  \\ 
0.00168  &   1.58648   & 13.4 $\pm$ 0.7 & \cite{Balewski1998}   \\ 
0.00268  &   1.58920   & 36.6 $\pm$ 2.6 & \cite{Balewski1998}   \\ 
0.00368  &   1.59192  &  63.0 $\pm$ 3.1  & \cite{Balewski1998}  \\ 
0.00468  &   1.59464 & 92.2 $\pm$  6.5  & \cite{Balewski1998}  \\ 
0.00568  &   1.59736  & 135 $\pm$ 11  & \cite{Balewski1998}   \\ 
0.00668  &   1.60008 & 164 $\pm$ 10   & \cite{Balewski1998}  \\  \hline 
0.0139  &   1.62000   & 630 $\pm$ 79  & \cite{Kowina2004}  \\ 
0.0159  &   1.62546  & 727 $\pm$ 57    & \cite{Kowina2004} \\ 
0.0202  &   1.63722  & 1011 $\pm$ 99   & \cite{Kowina2004}  \\ 
0.0301  &   1.66436  & 1366 $\pm$ 247   & \cite{Kowina2004}  \\ 
0.0397  &   1.69079  & 2118 $\pm$ 266   & \cite{Kowina2004}  \\ 
0.0593  &   1.74504  & 3838 $\pm$ 624   & \cite{Kowina2004}    \\  \hline 
\end{tabular}
\caption{ Total cross sections 
$\sigma_{tot} (pp \to pK \Lambda )$  from \cite{Balewski1998} and 
\cite{Kowina2004}. 
$Q = W -m_p-m_\Lambda - m_K$ is the excess energy.}
\l{tab:tb_vs_q_pp}
\end{table}
Let us dwell on some theoretical models of the $pp \to K^+\Lambda  p$
reaction. In the paper \cite{Laget1991}
the parameters of the model presenting the amplitude  
as a coherent sum
\be
M=M_{OPE}+M_{OKE}+M_{FSI(\pi )}+M_{FSI(K)}\;,
\l{modelamp}
\ee
had been chosen in order to match the existing total cross 
sections in the energy interval $2 < T_p <6$ GeV 
\cite{Baldini1988} \footnote{ 
In the next Laget's paper \cite{Laget2001} 
the data close to threshold, 
($ 0.68 \leq Q \leq 6.68$ MeV \cite{Balewski1998}) are already 
taken into account.}.
It had been shown, that the main contribution to the
total cross sections in this energy interval  
provides the one kaon exchange mechanism (OKE). Final state
interaction (FSI) contribute very little except the region
close to the kaon production threshold.
The Juelich group \cite{Gasparian2000} also
has stated that, in principle, $K$ exchange alone
could explain the total cross section especially
after inclusion of FSI effects. However the 
$\pi^0$ exchange cannot be neglected near the threshold. 
It was argued that the experimental data require
a destructive interference between 
$\pi^0$ and $K$ exchange contributions.
In the paper \cite{Sibirtsev2005} the detailed study
of the chances for identifying the reaction mechanism
of strangeness production are performed by considering the 
$\pi^0$ and $K$ exchange diagrams.

The aim of present work is to perform the calculations
as accurate as possible using as the amplitude
a coherent sum (\ref{modelamp}), thus following
the papers of J.M.Laget \cite{Laget1991,Laget2001}.
The model, which describes fairly well the total 
cross section in the wide interval of the excess 
energy $Q = W -m_p-m_\Lambda - m_K$ from 0.7 to 500 MeV,
is planned to use for the analysis of the proton and
kaon spectra in $pp \to p \Lambda K$ reaction at COSY
energies \cite{Valdau2006}. 

 Although near the threshold the spin flip effects
are negligible, the special attention is paid to the
complete taking into account of the spin dependence
of the considered amplitudes. It allows to make
realistic predictions of the polarization observables
for intermediate energies, which will be published elsewhere.
The Stapp formalism of the $M$-functions is the most proper formalism 
in this case and it is shortly described in the Appendix A.

\section{The total cross section}

The five-fold cross section is equal to
\be
{\d^5 \sigma \o \d p_p \d \Omega _p \d \Omega _K}=
{p_p^2 p_K^2 \o (2 \pi )^5 8E_p} \,
{1 \o 4p_b m} \,
{1 \o \left| p_KE_\Lambda -p_\Lambda E_Kz_{\Lambda K} \right|}|M|^2 \cqq
\l{eq:d5sig}
\ee
where  $z_{\Lambda K} \equiv {\bf p}_K.{\bf p}_\Lambda /p_K p_\Lambda$. 
The amplitude $M$ has been also calculated  as a coherent sum (\ref{modelamp})
where the separate contributions in this sum are considered in details
in the next sections

From eq.(\ref{eq:d5sig}) the total cross section  reads as 
\be
\sigma_{tot}={1 \o (2\pi)^5} {1 \o 32mp_b}
\int \d^3p_p \int \d \Omega _K {p_K^2 \o E_p} 
{1 \o \left | p_KE_\Lambda -p_\Lambda E_Kz_{\Lambda K} \right |}
|M|^2 \dqq
\l{eq:sigtot}
\ee
Let ${\bf p}_b$ be the momentum of the 
beam proton, ${\bf p}_b$ being directed along the z-axis.  
The total 4-momentum and the squared invariant 
mass of the system are equal
\[p_{tot}=(E,{\bf p}_b)=(m_p+\sqrt{m_p^2+p_b^2},{\bf p}_b)\;,\;\;
s=W^2=p_{tot}^2\;.\]
The proton momenta are placed then inside the ellipsoid
\begin{equation}
\frac{p_{px}^2+p_{py}^2}{q_{cm}^2}+
\frac{(p_{pz}-p_0)^2}{(\frac{E}{W}q_{cm})^2}=1\;,
\end{equation}
where 
\begin{equation}
q_{cm}^2=\frac{(s-(m_p+m_\Lambda+m_K)^2)(s-(m_p-m_\Lambda-m_K)^2)}{4s}
\end{equation}
is squared c.m. momentum of proton in the case of minimal mass of 
$K \Lambda$ sub-system equal to $m_\Lambda+m_K$ and
\begin{equation}
p_0 \equiv p_b\frac{s+m_p^2-(m_\Lambda+m_K)^2}{2s}\;.
\end{equation}

It is natural to prepare the integral (\ref{eq:sigtot})  for 
numerical integration over $d^3p_p$ by choosing the elliptical 
coordinates 
\begin{eqnarray}                        
p_{px}=a\sinh u \sin \theta ' \cos \phi  \nonumber \\
p_{py}=a\sinh u \sin \theta ' \sin \phi   \\
p_{py}=p_0+a\cosh u \cos \theta ' \;, \nonumber
\end{eqnarray}
where \[a \equiv q_{cm} \frac{p_{tot}}{W} \;,\] and
\[0 \leq u \leq u_{max}\] with \[ \sinh u_{max}=\frac{W}{p_{tot}}\;\;,
\;\;\cosh u_{max}=\frac{E}{p_{tot}}\;.\]
The Jacobian of transition from Cartesian to elliptical coordinates 
is equal to \[\d^3p_p =a^3(\sinh^2 u+\sin^2 \theta ')\sinh u \sin 
\theta '\;\d u  \d \theta ' \d \phi\] or 
\begin{equation}
\d^3 p_p =a^3(\cosh^2 u-x'^2) \d \cosh u \d x ' \d \phi\;.
\l{eq:jak}
\end{equation} 
Note that the volume of this ellipsoid is equal to
\[ V=\frac{4\pi}{3}q_{cm}^3\frac{E}{W}\;.\] 
Indeed it is calculated with the Jacobian (\ref{eq:jak}):
\begin{eqnarray*}
V=\int \d^3p=a^3 \int_1^{\frac{E}{p_{tot}}}  \d \cosh u \int_{-1}^1 \d x
\int_0^{2\pi} \d \phi (\cosh^2 u-x^2) = \\
2\pi a^3 \int_1^{\frac{E}{p_{tot}}} \d \cosh u (2 \cosh^2-\frac{2}{3})=
\frac{4\pi}{3}q_{cm}^3\frac{E}{W}
\end{eqnarray*}

We replace the measure $\d \Omega_K$, where
the angles $\Omega=(\theta _K,\phi_K)$ of kaon are defined with respect to 
fixed frame, by $\d \Omega_{K (K \Lambda)}$ with
the kaon  polar angles defined with respect to momentum of $K \Lambda$ 
sub-system. In this case we can make the trivial integration over azimuthal
angle of proton and obtain from eq.(\ref{eq:sigtot})
\bea
\sigma_{tot}= {1 \o (2\pi)^4} {1 \o 32mp_b}  a^3 
\int_1^{E/p_{tot}} \d \cosh u \int_{-1}^1 dx' \int \d \Omega_{K (K \Lambda)} \nn 
(\cosh^2 u-x'^2) {p_K^2 \o E_p}  
{|M|^2 \o \left | p_KE_\Lambda -p_\Lambda E_Kz_{\Lambda K} \right |} \dqq
\l{sigtot1} 
\eea
Let the variables of integration in eq.(\ref{sigtot1}) be replaced by
variables $q_i,\;i=1,2,3,4$ such that $0 \le q_i \le 1$. Then
\bea
\cosh u=1+(\frac{E}{p_b}-1)q_1\;,&\;\; \d \cosh u=(\frac{E}{p_b}-1) \d q_1\;, \nn
x'=1+2q_2\;,&\;\;\d x'=2 \d q_2\;,  \nn
x_K=1+2q_3\;,&\;\;\d x_K=2 \d q_3\;,  \nn
\phi_K=2 \pi q_4\;,&\;\; \d \phi_K=2 \pi \d q_4 \dqq \nonumber
\eea
Then we can rewrite the eq.(\ref{sigtot1}) as
\bea
\sigma_{tot}= {1 \o (2\pi)^3} {a^3 \o 8mp_b} ( {E \o p_b}-1) 
\int_0^1 \d q_1 \int_0^1 \d q_2 \int_0^1 \d q_3  \int_0^1 \d q_4  \nn
(\cosh^2 u-x'^2)\frac{p_K^2}{E_p}  
{|M|^2 \o \left | p_KE_\Lambda -p_\Lambda E_Kz_{\Lambda K} \right |} \dqq
\l{sigtot2} 
\eea
 
With the eq.(\ref{sigtot2}) we have tested
the model considered in next sections by using the CERN
program RIWIAD, an adaptive multidimensional integration subroutine 
which permits numerical integration of a large class
of functions, in particular those that are irregular at the border of the
integration region.

\section{Pion exchange mechanisms}

In what follows the tensor-like notation for the amplitudes with
the indices characterizing spin  projections is used. The
upper and lower indices  relate to the final and initial channels,  
respectively. The same index at the up and down positions stands
for a summation over this index. So, the $pp \to p \Lambda K^+$
amplitude looks like
\[
M_{\sigma_t \sigma_b}^{ \sigma_p  \sigma_\Lambda} 
( p_K,p_\Lambda,p_p   ; p_t,p_b) \cqq
\]
where $\sigma_t, \sigma_b, \sigma_p$ and $\sigma_\Lambda$ are
the spin projections of the target, beam, final proton and $\Lambda$
respectively. $p_K,p_\Lambda,p_p,p_t,p_b$ are the corresponding momenta.

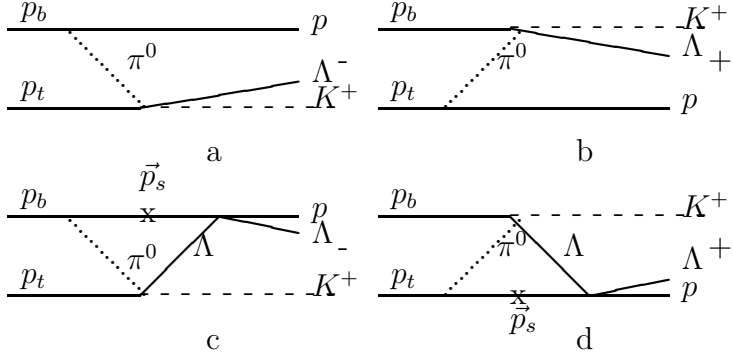
\begin{figure}[t]

\begin{picture}(320,70)(0,70)
\thicklines
\put(15,125){\line(1,0){110}} \put(20,130){$p_b$}
\put(130,125){$p$}
\put(15,95){\line(1,0){50}} \put(20,100){$p_t$}
\put(65,95){\line(6,1){60}} \put(130,105){$\Lambda$}
\put(65,93){- - - - - - - - - -}\put(130,95){$K^+$} 
\multiput(37,123)(2,-2){15}{.}
\put(60,110){$\pi^0$}
\put(140,110){-}  
\put(90,75){a}    
  
\put(155,95){\line(1,0){110}} \put(160,100){$p_t$} 
\put(270,95){$p$}
\put(155,125){\line(1,0){50}} \put(160,130){$p_b$}
\put(205,125){\line(6,-1){60}} \put(270,115){$\Lambda$}
\put(205,123){- - - - - - - - - -}\put(270,125){$K^+$} 
\multiput(207,123)(-2,-2){15}{.}
\put(200,110){$\pi^0$}
\put(280,110){+}  
\put(230,75){b}    
\end{picture}

\begin{picture}(320,70)(0,70)
\thicklines
\put(15,125){\line(1,0){110}} \put(20,130){$p_b$}
\put(65,137){$\vec p_s$} \put(65,123){x}
\put(85,110){$\Lambda$}
\put(15,95){\line(1,0){50}} \put(20,100){$p_t$}
\put(65,95){\line(1,1){30}} \put(130,115){$\Lambda$}
\put(95,125){\line(5,-1){30}} \put(130,125){$p$}
\put(65,93){- - - - - - - - - -}\put(130,95){$K^+$} 
\multiput(37,123)(2,-2){15}{.}
\put(60,105){$\pi^0$}
\put(140,110){-}  
\put(90,75){c}    
\put(155,95){\line(1,0){110}} \put(160,100){$p_t$} 
\put(205,83){$\vec p_s$} \put(205,92){x}
\put(225,110){$\Lambda$}
\put(155,125){\line(1,0){50}} \put(160,130){$p_b$}
\put(205,125){\line(1,-1){30}} \put(270,105){$\Lambda$}
\put(235,95){\line(5,1){30}} \put(270,95){$p$}
\put(205,123){- - - - - - - - - -}\put(270,125){$K^+$} 
\multiput(207,123)(-2,-2){15}{.}
\put(200,110){$\pi^0$}
\put(280,110){+}  
\put(230,75){d}    
\end{picture}
\caption{OPE (a,b) and FSI diagrams (c,d)}
\label{fig:OPE}
\end{figure}

\subsection{One Pion Exchange amplitude}

The one pion exchange(OPE) graphs are shown in Figs. \ref{fig:OPE} (a,b).
Let us start with the contribution of the graph $a$:
\be
i{{M_{(\pi^0 p \to K^+ \Lambda)}}_{\sigma_t}^{\sigma_\Lambda}
(p_\Lambda,p_K;p_t,p_b-p_p) 
{G_{(\pi^0 pp)}}^{\sigma_p}_{\sigma_b}(p_p,p_b-p_p;p_b) \o
t-m_\pi^2} \cqq 
\l{eq:OMEa}
\ee
where $t=(p_b-p_p)^2<0$ is the squared mass of the virtual $\pi^0$,
$M_{(\pi^0 p \to K^+ \Lambda)}$ is 
the half-off-shell $\pi^0 p \to K^+  \Lambda$ amplitude and 
$G$ is the $\pi^0 pp$ vertex which is equal in Stapp formalism to
(see Appendix A)
\be
\l{eq:piNN vertex}
{G_{(\pi^0 pp )}}^{\sigma_p}_{\sigma _b}(p_p,p_b-p_p;p_b)=
{g_\pi \o \sqrt{2}}f(t)       
m (e- \tilde{V}_p V_b)^{\sigma_p}_{\sigma _b} \; ,
\ee
where $g_{\pi}$ is the $\pi NN$ coupling constant, and 
$f(t)$ is the pion form factor for which we take the monopole 
representation 
\be
\l{eq:piNN form factor} 
f(t) ={m_\pi ^2-\Lambda_\pi^2 \o t-\Lambda_\pi ^2}  \cqq
\ee
$\Lambda_\pi$ being the cut-off parameter. 
Defining the pion 'wave function' in the nucleon
\be
{\Phi_{(\pi^0 pp)}}^{\sigma_p}_{\sigma _b} \equiv
{{G_{(\pi^0 pp)}}^{\sigma_p}_{\sigma _b} \o t-m_\pi^2}=   \\
{m g_\pi \o \sqrt{2}}\left( {1 \o t-m_\pi^2}-
{1 \o t-\Lambda_\pi ^2} \right)
(e- \tilde{V}_p V_b)^{\sigma_p}_{\sigma _b} \cqq
\l{eq:piNNwf}           
\ee
and replacing the amplitude $M_{(\pi^0 p \to K^+ \Lambda )}$
by the explored in the Rutherford Laboratory amplitude 
$M_{(\pi^- p \to  K^0 \Lambda)}$
\cite{Rutherford_Laboratory}, which are connected as follows
\[M_{(\pi^0 p \to K^+ \Lambda)}=
{M_{(\pi^- p \to K^0 \Lambda})}{\sqrt{2}} \cqq\]
we rewrite the eq.(\ref{eq:OMEa}) as
\be
{i \o \sqrt{2}}
{M_{(\pi^- p \to K^0 \Lambda )}}_{\sigma_t}^{\sigma_\Lambda}
(p_\Lambda,p_K;p_t,p_b-p_p) 
{\Phi_{(\pi^0 pp)}}^{\sigma_p}_{\sigma _b}(p_p,p_b-p_p;p_b) \dqq
\l{eq:OMEa_wf}
\ee
The contribution of the graph  $b$ pictured in Fig. \ref{fig:OKE}  
is obtained from the expression (\ref{eq:OMEa_wf}) by the interchange
of the beam and target quantities and we derive the final 
expression of the OPE contribution
\bea
{M_{(OPE)}}^{\sigma_\Lambda \sigma_p}_
{\sigma _b \sigma _t}(p_p,p_\Lambda,p_K;p_b,p_t)=  \hspace{2cm}  \nn
{i \o \sqrt{2}}
{M_{(\pi^- p \to K^0 \Lambda)}}_{\sigma_t}^{\sigma_\Lambda}
(p_\Lambda,p_K;p_t,p_b-p_p) 
{\Phi_{(\pi^0 pp)}}^{\sigma_p}_{\sigma _b}(p_p,p_b-p_p;p_b)- \nn
(b \leftrightarrow t) \cqq    
\l{eq:OPEcontribution} 
\eea
where the relative minus sign is required by Pauli principle.
In the Appendix A it is explained how the M-function expression 
(\ref{eq:piNN vertex}) of the $\pi^0 pp$ vertex is derived and how to obtain 
the M-function of the reaction $\pi^- p \rightarrow K^0 \Lambda$ from 
the canonical  flip and non-flip complex amplitudes $G^s$ and $H^s$. 
We have taken into account only the S- and P-wave partial amplitudes in the 
partial wave development\cite{Hohler1979}
\bea
G^s={1 \o q}\sum_{l=0}^{\infty} \left[(l+1)T_{l+}+lT_{l-}\right]P_l(x) \cqq \nn
H^s={1 \o q}\sum_{l=1}^{\infty} \left[T_{l+}-T_{l-}\right]P^1_l(x) \cqq
\l{Gs,Hs}
\eea
where $q=\sqrt{q_i q_f}$, $q_{i,f}$ are the c.m. momenta in initial and final 
channels and $x$ is the cosine of the c.m. scattering angle.
Following the papers \cite{Rutherford_Laboratory}(Baker1978) we took
\begin{eqnarray*}
T_{0+}(s)=\frac{\Gamma \sqrt{xx^\prime} m_p}{s-m_r^2-im_r\Gamma}+
\frac{2aq}{\sqrt{s}}\;   \\
T_{1\pm}(s)=\frac{\Gamma \sqrt{xx^\prime} e^{i\Phi} m_p}{s-m_r^2-im_r\Gamma}\;,
\end{eqnarray*}
where
\begin{equation}
\Gamma=\Gamma_r \left(\frac{q}{q_r}\right)^{2l+1}
\frac{2m_r}{\sqrt{s}+m_r} \frac{1+(1.7q_r)^2}{1+(1.7q)^2}
\end{equation}
and the parameters are given in  Tab. \ref{tab:Baker}.
Unfortunately the complex coefficient $a$ describing the background is not
presented by Rutherford Laboratory group, so we were to fit it to describe
the differential cross sections at four c.m. energies 1.661, 1.683, 1.694
and 1.724 GeV given in \cite{Rutherford_Laboratory}(Baker1978).
\begin{table}[t]
\caption{Resonance parameters of the 
$\pi^- p \rightarrow K^0 \Lambda$ amplitude}
\l{tab:Baker}
\centering
\begin{tabular}{|c|c|c|c|c|} 
   \hline 
         &  $m_r$  & $\Gamma_r$ & $\sqrt{xx^\prime}$ & $\Phi$   \\  \hline
$S_{11}$ &  1.68   & 0.09       & -0.25       & $0^0$  \\  
$P_{11}$ &  1.67   & 0.15       & -0.115      & $70^0$ \\  
$P_{13}$ &  1.75   & 0.40       & -0.09       & $20^0$ \\  \hline
\end{tabular}
\end{table}

The {\em virtuallity of the initial $\pi^0$} is taken into account as follows.
At first the amplitudes (\ref{Gs,Hs}) are multiplied by the
same form factor (\ref{eq:piNN form factor}) as the $\pi^0 pp$ vertex
(\ref{eq:piNN vertex}) 
\bea
G^s={f(t) \o \sqrt{q_i q_f}}
\sum_{l=0}^{\infty} \left[(l+1)T_{l+}+lT_{l-}\right]P_l(x) \cqq \nn
H^s={f(t) \o \sqrt{q_i q_f}}
\sum_{l=1}^{\infty} \left[T_{l+}-T_{l-}\right]P^1_l(x) \dqq
\l{fGs,fHs}
\eea
In addition the c.m. momentum in initial channel $\pi^0 p$ is calculated
with the virtual mass of the pion $t=(p_b-p_p)^2<0$:
\be
q_i={ \sqrt{[(W_{K \Lambda}-m_p)^2-t]
[(W_{K \Lambda}+m_p)^2-t]} \o 2 W_{K \Lambda}} \cqq
\l{qcm_pi0}
\ee
where $W_{K \Lambda}$ is the invariant mass of the reaction.

\subsection{FSI mechanism with the intermediate pion}

The FSI graphs are shown in Figs. \ref{fig:OPE} $c,d$. Writing the
contribution of, for example, the graph $c$ in accordance with Feynman rules 
and putting the proton 'spectator' $p_s$ on the mass-shell  we obtain
\footnote{As it was explained above, the same index $\sigma_v$, 
spin projection of the virtual $\Lambda$, at the up and down positions stands
for a summation over this index.}
\bea
 M_{FSI(\pi)c}{}_{\sigma_t \sigma_b}^{ \sigma_p  \sigma_\Lambda} =-i \int
{\d {\bf p}_ s \o (2\pi)^3 2E_s} \hspace{2cm} \nn
M_{(p_s\Lambda_v \to p\Lambda)}{}_{\sigma_v}^{\sigma_\Lambda}
{1 \o m^2_v-m^2_\Lambda+i\epsilon }           
M_{(\pi^0 p_t \to K^+ \Lambda_v)}{}_{\sigma_t}^{\sigma_v}  
\Phi_{(\pi^0 pp)}{}_{\sigma_b}^{\sigma_s}  \approx \nn
-i {m_p \o W_{p\Lambda}} \int
{\d {\bf p}_ s \o (2\pi)^3 2E_s} \hspace{2cm} \nn
M_{(p_s\Lambda_v \to p\Lambda)}{}_{\sigma_v}^{\sigma_\Lambda}
{1 \o q_{cm}^2-\xi^2+i\epsilon }           
M_{(\pi^0 p_t \to K^+ \Lambda_v)}{}_{\sigma_t}^{\sigma_v}  
\Phi_{(\pi^0 pp)}{}_{\sigma_b}^{\sigma_s} \cqq
\l{eq:FSIc} 
\eea
where  $m_v$ is the mass of virtual $\Lambda$ inside the triangle loop,
the pion 'wave function' in the nucleon $\Phi_{(\pi^0 pp)}$ is defined
in eq. (\ref{eq:piNNwf}) and the  non-relativistic approximation
of the $\Lambda$ propagator 
\[m_v^2-m_\Lambda ^2 \approx 
{W_{p\Lambda} \o m}(q_{cm}^2-\xi^2)\]
is used. Here
$W_{p\Lambda}=\sqrt{s_{p\Lambda}}$, $s_{p\Lambda}=p_{p\Lambda}^2$ and
\bea
q_{cm}={\sqrt{[(W_{p\Lambda}-m_p)^2-m_\Lambda^2]
[(W_{p\Lambda}+m_p)^2-m_\Lambda^2]} \o 2W_{p\Lambda}} \cqq \nn
\xi = {\sqrt{[(W_{p\Lambda}-m_p)^2-m_v^2]
[(W_{p\Lambda}+m_p)^2-m_v^2]} \o 2W_{p\Lambda}} \qq  
\eea
are the c.m. momenta of $\Lambda$ after (on-shell) and before
(off-shell) the scattering.

Simplifications are necessary in order to
calculate the integral (\ref{eq:FSIc}) since it requires 
a knowledge of the  off-shell $\pi^0 p \to K^+ \Lambda$ and  $p\Lambda$  
amplitudes. In this we have followed the paper
\cite{Laget1978} with some modification described in \cite{Grebenyuk2005}.
Namely, these amplitudes are taken out of the integral sign at some momentum  
$p_s^0$ placed on the singular surface of the integral  corresponding to 
the mass-shell of the virtual $\Lambda$.

When a separable potentials for $p\Lambda$ scattering are assumed
for each partial-wave state the relation
\be
M^{off}_{p\Lambda}=f(m^2_v)M^{on}_{p\Lambda} \qq
\l{eq:offshellamp}
\ee
is strictly valid. The usual form of the form factor $f$ is
\be
f(m^2_v)={q^2_{c.m.}+\beta^2 \o \xi^2+\beta^2} \cqq
\l{eq:Yamaguchi}
\ee
where $\beta$ is the cut-off parameter.
Near threshold  two S-waves, $^1S_0$ and $^3S_1$,
give the major contribution. In the paper\cite{Takahashi1980}  
different sets of separable $p\Lambda$ potential were found with
good fits to the cross-section. They are presented in Tab. \ref{tab:beta}.
Following this table we  neglected the difference between the spin
singlet and spin triplet S-wave parameters and  took the form factor
(\ref{eq:Yamaguchi}) for the whole low-energy amplitude  with $\beta$
varying in the region $0.1 \leq \beta \leq 0.3$ GeV/c.
\begin{table}[t]
\centering
\begin{tabular}{|c|c|c|c|c|c|} 
   \hline 
        &  Set A-1 & Set A-2 & Set A-3 & Set B-1  & Set B-2 \\  \hline
$^1S_0$ &  0.209   & 0.256   & 0.239   & 0.257    &  0.258    \\  
$^3S_1$ &  0.217   & 0.246   & 0.252   & 0.247    &  0.243   \\   \hline
\end{tabular}
\caption{Parameter $\beta$(GeV/c) of separable $p\Lambda$ potential 
by Y.Takahashi et al.} 
\l{tab:beta}
\end{table}
Taking into account that
\[
{f(m^2_v) \o q_{cm}^2-\xi^2+i\epsilon } =
\left[-{1 \o \xi^2-q_{cm}^2-i\epsilon}+{1 \o \xi^2+\beta^2}\right] \qq
\]
the eq. (\ref{eq:FSIc}) can be rewritten as 
\newpage
\bea
 M_{FSI(\pi)c}{}_{\sigma_t \sigma_b}^{\sigma_p \sigma_\Lambda} \approx 
-i {m_p \o W_{p\Lambda}} \int {\d {\bf p}_ s \o (2\pi)^3 2E_s} \hspace{2cm} \nn  
M^{on}_{(p_s\Lambda_v \to p\Lambda)}{}_{\sigma_v}^{\sigma_\Lambda}
M_{(\pi^0 p_t \to K^+ \Lambda_v)}{}_{\sigma_t}^{\sigma_v}  
\Phi_{(\pi^0 pp)}{}_{\sigma_b}^{\sigma_s} 
\left[-{1 \o \xi^2-q_{cm}^2-i\epsilon}+{1 \o \xi^2+\beta^2}\right] \dqq
\l{eq:FSIc1} 
\eea

We  took the half-off-shell $\pi^0 p \to K^+\Lambda$ and on-shell part of 
$p\Lambda$ amplitudes (\ref{eq:offshellamp}) out of the 
integral sign at specially chosen momenta. Following the argumentation of 
\cite{Grebenyuk2005} the best choice for the graph $c$ is 
the maximal momentum of the on-shell loop proton $p_s$, which is 
directed along ${\bf p}_{\Lambda}$ and is equal to
\[
{\bf p}_{p(+)}={\bf p}_{p\Lambda}( {E^{cm}_p \o W_{p\Lambda}} + 
{E_{p\Lambda} \o W_{p\Lambda}}
{q^{cm} \o p_{p\Lambda}}) \dqq
\]
 In this case $\Lambda$ is on-shell 
too and has correspondingly  the minimal momentum $p_{\Lambda (-)}$
\[
{\bf p}_{\Lambda (-)}=
{\bf p}_{p\Lambda}({E^{cm}_\Lambda \o W_{p\Lambda}} - 
{E_{p\Lambda} \o W_{p\Lambda}}\;  
{q^{cm} \o p_{p\Lambda}}) \dqq   
\]
Here $E^{cm}_p$ and $E^{cm}_\Lambda$ are c.m. energies of the proton
and $\Lambda$ respectively.
Under these assumptions we can rewrite eq.(\ref{eq:FSIc1}) 
as follows 
\newpage   
\bea
{ M_{FSI(\pi)c}}^{\sigma_\Lambda \sigma_p}_
{\sigma_b \sigma_t}(p_p,p_\Lambda,p_k;p_b,p_t)=  \hspace{2cm}   \nn
-i{M_{(p\Lambda )}}^{\sigma_p \sigma_\Lambda}_{\sigma_s \sigma_v}         
(p_p,p_\Lambda ;p_{p(+)},p_{\Lambda (-)}) 
(e- \tilde{V}_{p(+)} V_b)^{\sigma_s}_{\sigma_b} \nn 
{ M_{(\pi^0 p \to K^+\Lambda)}}_{\sigma_t}^{\sigma_v} 
(p_K,p_{\Lambda (-)};p_b-p_{p(+)},p_t) \cdot F(s_{p\Lambda},t)
  \cqq   \hspace{2cm}
\eea
where $t=(p_t-p_K)^2$ and
\bea
F(s_{p\Lambda},t) \equiv {g_{\pi}m^2 \o  W_{p\Lambda} \sqrt{2} } 
\int {\d {\bf p}_ s \o (2\pi)^3 2E_s}\;\left[
{1 \o (p_b-p_s)^2-m_{\pi}^2}-{1 \o (p_b-p_s)^2-\Lambda_{\pi}^2}\right] \nn
\left[-{1 \o \xi^2-q_{cm}^2-i\epsilon}+{1 \o \xi^2+\beta^2}\right] \dqq
\l{eq:secondFint}
\end{eqnarray}
In Appendix B the details of the calculation of the integral (\ref{eq:secondFint})
are given. The result is 
\bea
F(s_{p\Lambda},t) \approx {g_{\pi} m \o 32\pi W_{p\Lambda} q_b \sqrt{2}} 
\sum_j C_j  \hspace{2cm}   \nn
\left[-i\ln {\xi^2_{j+}+\alpha^2_j \o \xi^2_{j-}+\alpha^2_j}-
2\arctan {\xi_{j+} \o \alpha_j}-2\arctan {\xi_{j-} \o \alpha_j}+
4\arctan {\xi_{j+}+\xi_{j-} \o 2(\beta+\alpha_j)}\right] \cqq
\eea
where $q_b$ is the c.m. momentum of the system $p_b+(p_t-p_K)$ equal to
\be
q_b={\sqrt{[(W_{p\Lambda}+m)^2-t][(W_{p\Lambda}-m)^2-t]} \o 2W_{p\Lambda}} \cqq
t=(p_t-p_K)^2
\ee
and
\bea
\alpha_1=E_b\sqrt{1-(1-{m_{\pi}^2 \o 2m^2})^2} \cqq & \qq
\alpha_2=E_b\sqrt{1-(1-{\Lambda_{\pi}^2 \o 2m^2})^2} \cqq \nn
\xi_{1\pm}=q_b(1-{m_{\pi}^2 \o 2m^2})\pm q_{c.m.} \cqq & \qq
\xi_{2\pm}=q_b(1-{\Lambda_{\pi}^2 \o 2m^2})\pm q_{c.m.} \cqq  \\
C_1=1 \cqq & \qq C_2=-1 \dqq  
\eea

The contribution of the second FSI graph with virtual pion 
(Fig. \ref{fig:OPE} $d$) is  derived by the interchange
$ b \leftrightarrow t$:
\bea
{ M_{FSI(\pi)d}}^{\sigma_\Lambda \sigma_p}_
{\sigma_b \sigma_t}(p_p,p_\Lambda,p_k;p_b,p_t)=  \hspace{2cm}   \nn
-i{M_{(p\Lambda )}}^{\sigma_p \sigma_\Lambda}_{\sigma_s \sigma_v}         
(p_p,p_\Lambda ;p_{p(-)},p_{\Lambda (+)}) 
(e- \tilde{V}_{p(-)} V_t)^{\sigma_s}_{\sigma_b} \nn 
{ M_{(\pi^0 p \to K^+\Lambda)}}_{\sigma_t}^{\sigma_v} 
(p_K,p_{\Lambda (+)};p_t-p_{p(-)},p_b) \cdot F(s_{p\Lambda},t)
  \cqq   \hspace{2cm}
\end{eqnarray}
where $t=(p_K-p_b)^2$, and
\bea
{\bf p}_{p(-)}={\bf p}_{p\Lambda}( {E^{cm}_p \o W_{p\Lambda}} - 
{E_{p\Lambda} \o W_{p\Lambda}}
{q^{cm} \o p_{p\Lambda}}) \cqq 
{\bf p}_{\Lambda (+)}=
{\bf p}_{p\Lambda}({E^{cm}_\Lambda \o W_{p\Lambda}} - 
{E_{p\Lambda} \o W_{p\Lambda}}\;  
{q^{cm} \o p_{p\Lambda}}) \dqq   
\eea

\section{Kaon exchange mechanisms}

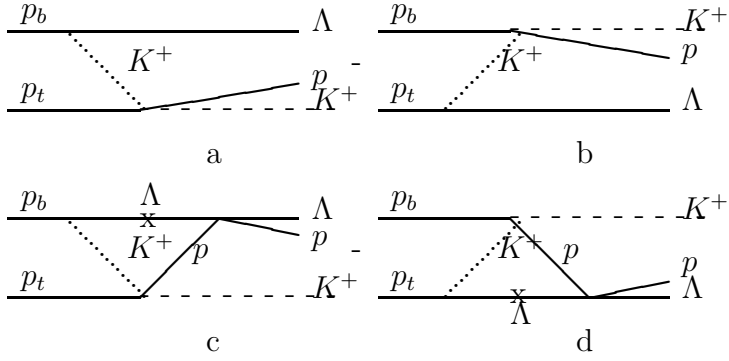
\begin{figure}[t]

\begin{picture}(320,70)(0,70)
\thicklines
\put(15,125){\line(1,0){110}} \put(20,130){$p_b$}
\put(130,125){$\Lambda$}
\put(15,95){\line(1,0){50}} \put(20,100){$p_t$}
\put(65,95){\line(6,1){60}} \put(130,105){$p$}
\put(65,93){- - - - - - - - - -}\put(130,95){$K^+$} 
\multiput(37,123)(2,-2){15}{.}
\put(60,110){$K^+$}
\put(145,110){-}  
\put(90,75){a}    
\put(155,95){\line(1,0){110}} \put(160,100){$p_t$} 
\put(270,95){$\Lambda$}
\put(155,125){\line(1,0){50}} \put(160,130){$p_b$}
\put(205,125){\line(6,-1){60}} \put(270,115){$p$}
\put(205,123){- - - - - - - - - -}\put(270,125){$K^+$} 
\multiput(207,123)(-2,-2){15}{.}
\put(200,110){$K^+$}
\put(230,75){b}   
\end{picture}

\begin{picture}(320,70)(0,70)
\thicklines
\put(15,125){\line(1,0){110}} \put(20,130){$p_b$}
\put(65,130){$\Lambda$}  \put(65,122){x}
\put(85,110){$p$}
\put(15,95){\line(1,0){50}} \put(20,100){$p_t$}
\put(65,95){\line(1,1){30}} \put(130,115){$p$}
\put(95,125){\line(5,-1){30}} \put(130,125){$\Lambda$}
\put(65,93){- - - - - - - - - -}\put(130,95){$K^+$} 
\multiput(37,123)(2,-2){15}{.}
\put(60,110){$K^+$}
\put(145,110){-}  
\put(90,75){c}    
\put(155,95){\line(1,0){110}} \put(160,100){$p_t$} 
\put(205,85){$\Lambda$} \put(205,93){x}
\put(225,110){$p$}
\put(155,125){\line(1,0){50}} \put(160,130){$p_b$}
\put(205,125){\line(1,-1){30}} \put(270,105){$p$}
\put(235,95){\line(5,1){30}} \put(270,95){$\Lambda$}
\put(205,123){- - - - - - - - - -}\put(270,125){$K^+$} 
\multiput(207,123)(-2,-2){15}{.}
\put(200,110){$K^+$}
\put(230,75){d}    
\end{picture}
\caption{OKE (a,b) and FSI diagrams (c,d)}
\label{fig:OKE}
\end{figure}

\subsection{One Kaon Exchange amplitude}

The OKE graphs are shown in Figures \ref{fig:OKE}a,b. The final 
expression of the OKE contribution is similar to eq.(\ref{eq:OPEcontribution})
\newpage
\bea
{M_{(OKE)}}^{\sigma_\Lambda \sigma_p}_
{\sigma _b \sigma _t}(p_p,p_\Lambda,p_K;p_b,p_t)=   \hspace{2cm}  \nn
i{M_{(K^+ p)}}_{\sigma_t}^{\sigma_p}(p_p,p_K;p_t,p_b-p_\Lambda) 
{\Phi_{(K^+p\Lambda)}}^{\sigma_\Lambda}_{\sigma _b}(p_\Lambda,p_b-p_\Lambda;p_b)-  
(b \leftrightarrow t) \dqq
\label{eq:OKEcontribution} 
\eea
The kaon 'wave function' in the $K^+p\Lambda$ vertex is equal to
\be
{\Phi_{(K^+p\Lambda )}}^{\sigma_\Lambda}_{\sigma _b}=
g_K \sqrt{m m_\Lambda} \left( {1 \o t-m_K^2}-
{1 \o t-\Lambda_K ^2} \right)
(e- \tilde{V}_\Lambda V_b)^{\sigma_\Lambda}_{\sigma _b} \cqq
\l{eq:KNLwf}                   
\ee
where $t=(p_\Lambda-p_b)^2$, 
$g_K$ is the  $K^+p\Lambda$ coupling constant \cite{Adelseck1990} and
$\Lambda_K$ is the cut-off parameter in the monopole form factor
(see eq. (\ref{eq:piNN form factor}) for  virtual pion) of virtual $K$.

The $M_{(K^+ p)}$ is
the half-off-shell $K^+ p \to K^+ p$ amplitude, the M-function of which
is derived from the canonical amplitude in the same way as 
$M_{(\pi^- p \to K^0 \Lambda})$ (see the previous section and Appendix A).
In order to calculate the canonical flip and non-flip amplitudes 
we have utilized the
isotriplet partial-wave amplitudes below 1.5 GeV/c obtained in the phase shift 
analysis by Martin \cite{Martin1975}.

\subsection{FSI with intermediate kaon}

For the graph in Fig. \ref{fig:OKE} $c$ we have
\bea
{ M_{FSI(K)c}}^{\sigma_\Lambda \sigma_p}_
{\sigma_b \sigma_t}(p_p,p_\Lambda,p_k;p_b,p_t)=   \hspace{2cm}  \nn
-i{M_{(p\Lambda)}}^{\sigma_p \sigma_\Lambda}_{\sigma_v \sigma_s}
(p_p,p_\Lambda ;p_{p(-)},p_{\Lambda (+)})\; 
(e- \tilde{V}_{\Lambda(+)} V_b)^{\sigma_s}_{\sigma_b} \nn 
{ M_{(K^+ p) }}_{\sigma_t}^{\sigma_v} 
(p_K,p_{p(-)};p_b-p_{\Lambda(+)},p_t) \cdot F_K(s_{p\Lambda},t) 
\cqq  \hspace{2cm} 
\eea
where $t=(p_K-p_t)^2$
and
\newpage
\bea
F(s_{p\Lambda},t) = {g_K  m_\Lambda \sqrt{m m_\Lambda} \o W_{p\Lambda}} 
\int { \d {\bf p}_s \o (2\pi)^3 2E_s}\;\left[
{1 \o (p_b-p_s)^2-m_K^2}-{1 \o (p_b-p_s)^2-\Lambda_K^2}\right] \nn
\left[-{1 \o \xi^2-q_{cm}^2-i\epsilon}+ {1 \o \xi^2+\beta^2}\right] \cqq
\l{eq:secondFintK}
\eea
where now the integration is performed over the 3-momentum ${\bf p}_s$ 
of the on-shell $\Lambda$ in the triangle loop. 
Manipulations similar to those  described in Appendix B for the pion case
result in
\bea
F_K(s_{p\Lambda},t) \approx {g_K \o 32\pi q} \sqrt{{m^3_\Lambda \o m^3}}
\sum_j C_j \nn
\left[-i\ln {\xi^2_{j+}+\alpha^2_j \o \xi^2_{j-}+\alpha^2_j}-
2\arctan {\xi_{j+} \o \alpha_j}-2\arctan {\xi_{j-} \o \alpha_j}+
4\arctan {\xi_{j+}+\xi_{j-} \o 2(\beta+\alpha_j)}\right] \cqq
\eea
where
\bea
\alpha_1=E_b\sqrt{1-{(m^2+m^2_\Lambda-m_K^2)^2 \o 4m^2m^2_\Lambda}} \cqq 
\alpha_2=E_b\sqrt{1-{(m^2+m^2_\Lambda-\Lambda_K^2)^2 \o 4m^2m^2_\Lambda}} \cqq \nn
\xi_{1\pm}=q_b{m^2+m^2_\Lambda-m_K^2 \o 2mm_\Lambda}\pm q_{c.m.} \cqq 
\xi_{2\pm}=q_b{m^2+m^2_\Lambda-\Lambda_K^2 \o 2mm_\Lambda}\pm q_{c.m.} \cqq  \\
C_1=1 \cqq  C_2=-1 \cqq  
\eea
and $q={W_{p\Lambda} \o m}q_b$.
The contribution of the second FSI graph with virtual kaon 
(Fig. \ref{fig:OKE} $d$) is  derived by the interchange
$ b \leftrightarrow t$.

\section{Results and discussion}

Let us start with the total cross sections 
for OKE+FSI mechanism. Free parameters in this case are
$g_K$ - $K^+p\Lambda$ coupling constant,
$\Lambda_K$ - the cut-off parameter in the monopole form factor
of virtual $K$ and $\beta$ - the cut-off parameter in the 
form factor in half-off shell  $p\Lambda$  interaction. 
The cut-off parameter $\Lambda_K=0.9$ had been chosen 
and with $g_K=-4.17\sqrt{4\pi}=-14.78$ GeV, the value recommended in the 
paper \cite{Adelseck1990}, it appeared that the experimental data
presented in Tab. \ref{tab:tb_vs_q_pp} are fairly well described by only
OKE mechanism. Still to find place for the FSI contributions,
from the one hand the  coupling constant $g_K$ had been minimally decreased
to the value $g_K=-11.8$ in order to stay in agreement with SU(3) predictions 
for the kaon-hyperon-nucleon coupling constants (see \cite{Adelseck1990}),
and from the other hand the FSI contributions had been minimized by
choosing rather small value of the half-off shell  $p\Lambda$ cut-off parameter: 
$\beta=0.1$. The resulting total cross sections versus 
the excess  energy $Q = W -m_p-m_\Lambda - m_K$
are shown as double logarithmic plot in Fig. \ref{fig:sigtot_oke}. 
\begin{figure}[ht]
\begin{center}
\epsfig{file=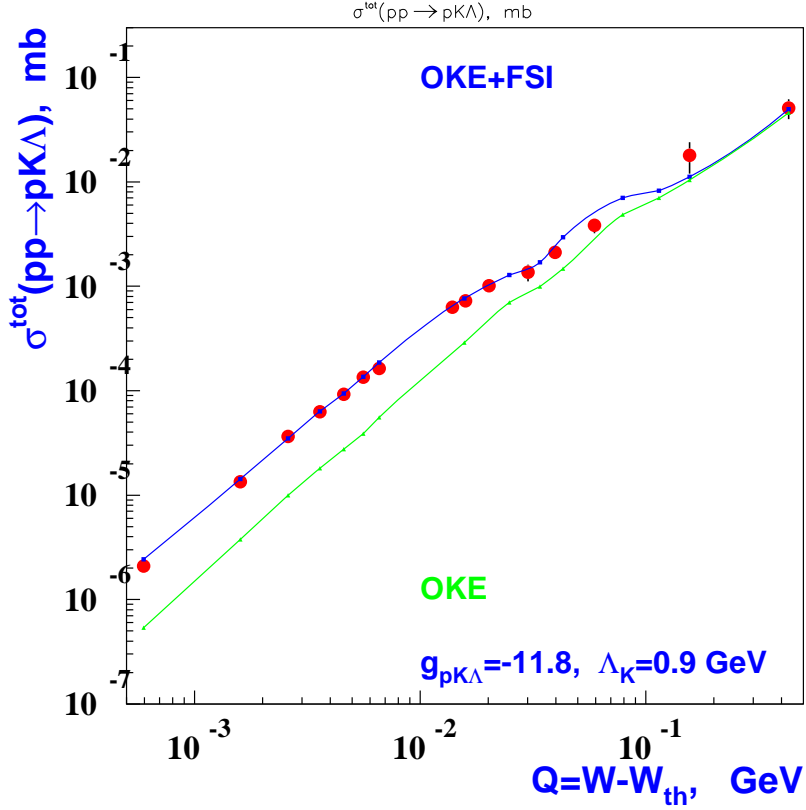,width=12cm}
\end{center}
\caption{ Red circles are the experimental total $pp \to p \Lambda K^+$
cross sections versus the excess  energy $Q = W - W_{th}$. 
Green curve presents the calculations in the frame of OKE model.
Blue curve - the calculations in the frame of OKE+FSI model.}
\l{fig:sigtot_oke}
\end{figure}
It is seen that OKE+FSI(K) model alone with the plausible parameters  
describes very well the experimental data, which is already a hint
in favor of kaon mechanism dominance in $pp \to p \Lambda K^+$
reaction. The $Q$ behavior of the difference between pure OKE and
OKE+FSI(K) curves demonstrates, that near the threshold FSI contribution
is compatible with the OKE one, whereas at higher Q the OKE becomes
the dominating mechanism.

Though the kaon mechanism seems to be enough we should find place
for the OPE+FSI($\pi$) graphs.  Free parameters in this case are
$g_{\pi}$ - the $\pi NN$ coupling constant and $\Lambda_\pi$ -
the cut-off parameter in the monopole form factor (\ref{eq:piNN form factor})
of virtual $\pi$.  The cut-off parameter in the 
form factor in half-off shell  $p\Lambda$  interaction  $\beta =0.1$ 
was fixed by OKE+FSI(K) fit. 
The well established value $g_{\pi} \simeq 13.5$ 
($g_{\pi}^2 / 4\pi = 14.5$)  had been fixed. Then it appeared that using
the common accepted $\Lambda_\pi \approx 1$ GeV results in high OPE
contribution near the threshold which would be impossible to compensate
by other mechanisms. We have decreased the cut-off parameter $\Lambda_\pi$
to the value $\Lambda_\pi = 0.515$ GeV. The resulting plots are shown
in Fig. \ref{fig:sigtot_ope}. 
\begin{figure}[ht]
\begin{center}
\epsfig{file=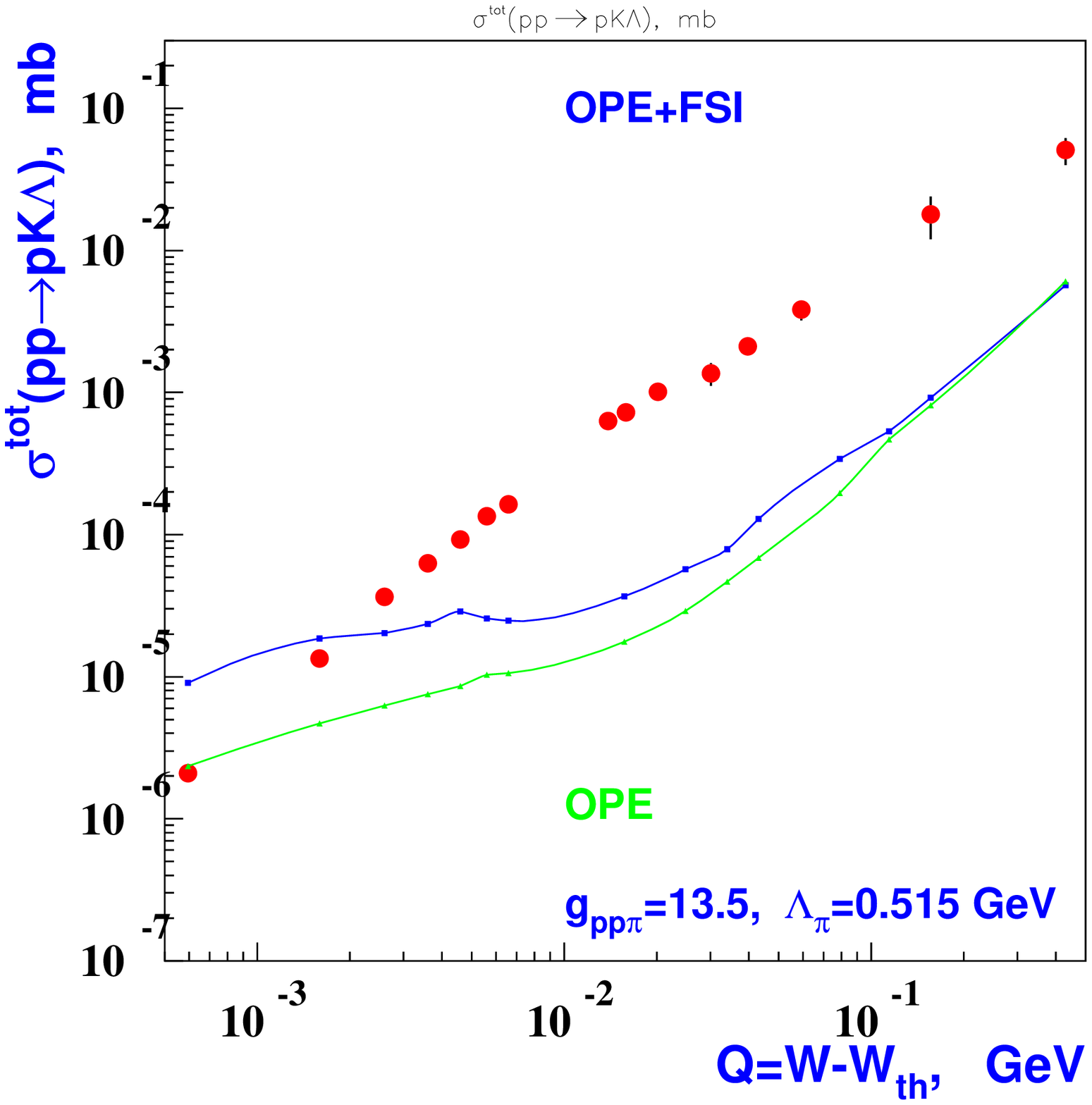,width=12cm}
\end{center}
\caption{ Red circles are the experimental total $pp \to p \Lambda K^+$
cross sections versus the excess  energy $Q = W - W_{th}$. 
Green curve presents the calculations in the frame of OPE model.
Blue curve - the calculations in the frame of OPE+FSI($\pi$) model.}
\l{fig:sigtot_ope}
\end{figure}
It is seen that only for $Q < 1.7$ MeV the contribution 
of OPE+FSI($\pi$) graphs exceeds experimental data and 
there is no possibility to decrease this contribution. 
For higher Q the OPE+FSI($\pi$) contribution to the
experimental data is negligible.

The applying the full model, including the both kaon and pion
mechanisms, with the parameters fixed above had shown that in case
of "constructive" interference
\[ M=M_{OPE}+M_{OKE}+M_{FSI(\pi )}+M_{FSI(K)} \]
the model total cross sections exceed the measured ones,
the difference being especially large near the threshold. In contrast,
the choice of destructive interference
\[ M=-M_{OPE}+M_{OKE}-M_{FSI(\pi )}+M_{FSI(K)} \]
results in fairly well description of the data, which is
demonstrated in Fig. \ref{fig:sigtot_oke-ope}.
\begin{figure}[ht]
\begin{center}
\epsfig{file=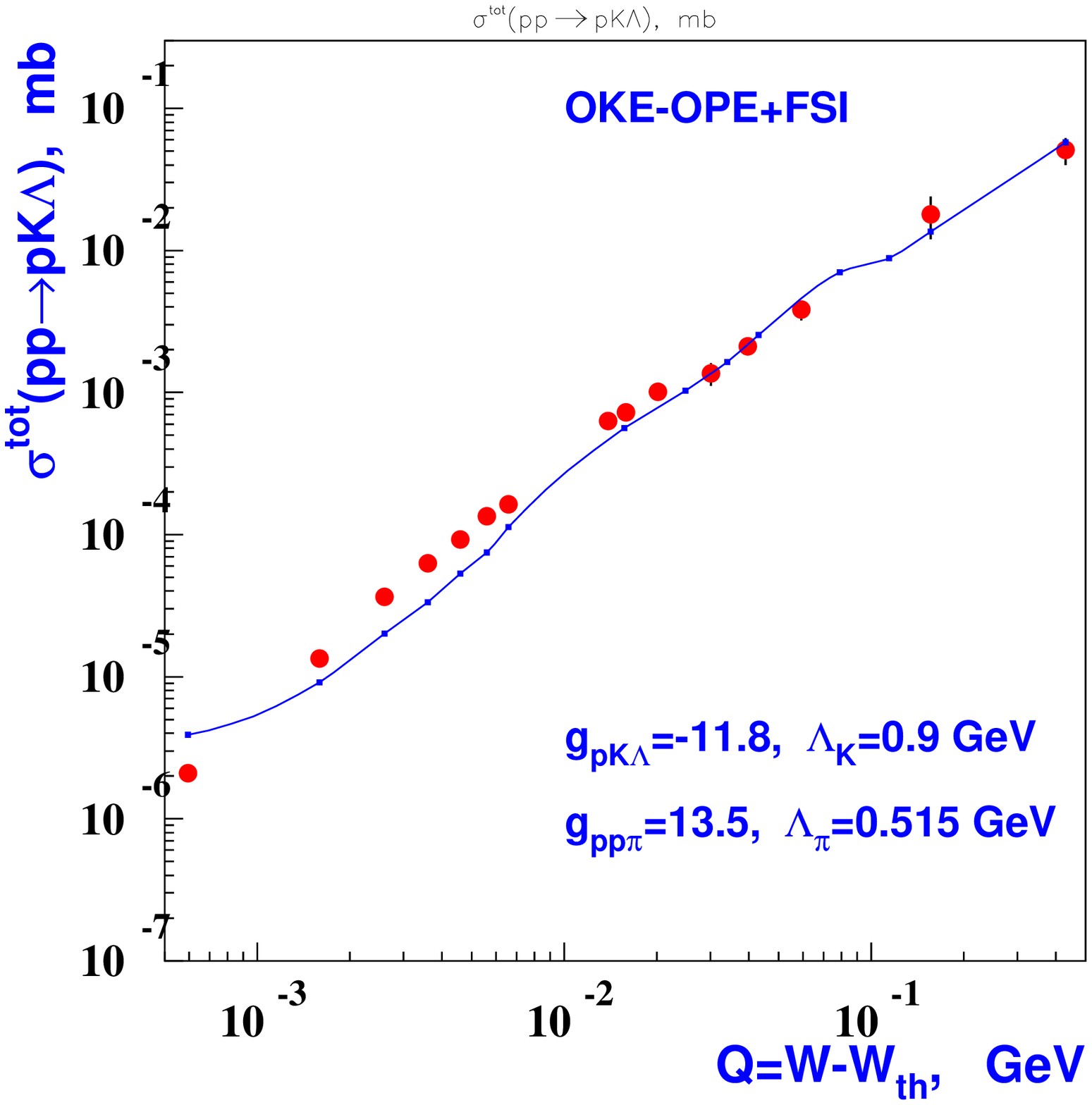,width=12cm}
\end{center}
\caption{Red circles are the experimental total $pp \to p \Lambda K^+$
cross sections versus the excess  energy $Q = W - W_{th}$. 
Blue curve - the calculations in the frame of full OKE-OPE model
(destructive interference)}.
\l{fig:sigtot_oke-ope}
\end{figure}
This our observation confirms the conclusion of
the Juelich group \cite{Gasparian2000} that
the experimental data require a destructive interference between 
$\pi^0$ and $K$ exchange contributions.

 In conclusion we can confirm
the main features of the $pp \to p \Lambda K^+$ scattering
at low and intermediate energies observed in the previous
studies \cite{Laget1991, Gasparian2000, Sibirtsev2005}:
the kaon mechanism dominance at intermediate energies,
the compatibility of all mechanisms near the threshold and
the destructive interference between 
kaon and pion amplitudes.  
 The model presented in this work describes sufficiently well the
measured total cross sections $pp \to p \Lambda K^+$
in the wide interval of the excess  energy 
\mbox{$Q = W - m-m_\Lambda-m_K$}:
$ 0.68 \leq Q \leq 430 $ MeV.
  Our further plans are to apply this model for the description
for the analysis of the proton and kaon spectra in 
$pp \to p \Lambda K$ reaction at COSY energies \cite{Valdau2006}
and to make predictions for the polarization observables
at these energies.

\section*{Acknowledgement}

The author thanks V.Koptev for the fruitful  discussions
and cooperation during performing of this work.

\appendix
\section{M-functions}

When calculating a coherent sums of various mechanisms
one needs in principle to transform the matrices of corresponding 
amplitudes from  center of mass frames to the common laboratory frame. 
These transformations depend on the spin basis chosen for the 
one-fermion states.  Usually the canonical or helicity basis are
used which are transformed with unitary two by two matrices depending on a 
fermion momentum. They are so called Wigner rotations. 
The covariant basis \cite{Stapp1983}transforming independently of the particle
momentum by  the  unimodular two by two  matrices  is
free from this deficiency. Amplitudes in the covariant basis are
usually referred to as the M-functions of Stapp. 
The covariant formalism of the M-functions
developed by Stapp \cite{Stapp1983} is based on the matrices
 $\sigma^\mu=(1, \vec{\sigma})$ and  $\tilde{\sigma}^\mu=(1, -\vec{\sigma})$, 
where $ \vec{\sigma}$ are the standard Pauli matrices. 
With each momentum 
 $P^\mu$ of the reaction  two by two matrices 
$\tilde{P} \equiv P^\mu \tilde{ \sigma_\mu}$ and $P \equiv P^\mu  \sigma_\mu$
are associated. The  products $\tilde{P}_i P_j$ of these matrices
are the  elements from which the M-functions are built.
The matrices  $\tilde{V}_i$ and $V_i$ associated with the
four-velocity of the i-th particle, serve as the metric tensors  
of the particle  when performing the contraction over 
this index (traces, successive processes).

Let us  explain the M-function expression (\ref{eq:piNN vertex}) of the 
$\pi^0 pp$ vertex and how to obtain the M-function of the reaction
$\pi^- p \rightarrow K^0 \Lambda$ from the canonical  amplitude.
As to the $\pi^0 pp$ vertex it is  nothing but the usual expression
$g \bar{u}_f \gamma_5 u_i$ with the exception that the modified Dirac
bispinors are used. In the Weyl representation they are equal to
\begin{equation}
{u^\alpha}_b=\sqrt{m}
\left(\begin{array}{c}
e^a_b \\
V_{\bar{a} b}
\end{array}\right) \;,\;\;\;
{\bar{u}^b}_\alpha=\sqrt{m}
\left(\begin{array}{cc}
e^b_a   & V^{b\bar{a}}
\end{array}\right)\;.
\label{eq:loc_mat}
\end{equation}
In the Weyl representation the $\gamma_5$-
matrix looks like \[\gamma_5 =\left(\begin{array}{cc} e & 0 \\
0 & -e \end{array}\right)\] and the amplitude of transition between real
states with the production of the pseudoscalar is equal to
\[ M^a_b(p_f,q;p_i)=
g{\bar{u}^a}_\alpha {\gamma_5}^\alpha _\beta {u^\beta}_b =
g\sqrt{m_fm_i}(e^a_b-V_f^{a\bar{c}}{V_i}_{\bar{c}b})\;,\]
or in indexless form
\begin{equation}
M(p_f,q;p_i)=g \bar{u}_f \gamma_5 u_i =
g\sqrt{m_fm_i}(e-\tilde{V}_f V_i)\;,
\label{eq:indless pseuds vertex}
\end{equation}
which explains the eq.(\ref{eq:piNN vertex}).

As to the M-function of the $0\frac{1}{2} \rightarrow 0\frac{1}{2}$
reaction it is possible to represent as
\begin{equation}
M=G^mb_1+H^mb_2 \;,
\label{eq:noninvMfunct}
\end{equation}
 where $G^m$ and $H^m$ are the complex functions
and $b_{1,2}$ are the basis M-functions equal to
\begin{equation}
b_1 \equiv
\frac{e+\tilde{V}_fV_i}{\sqrt{(M_1,M_1)}}\;,\;\;
b_2 \equiv
\frac{\alpha(e+\tilde{V}_f V_i)+(\tilde{V}_f V+\tilde{V} V_i)}
{\sqrt{(M_2,M_2)}}\;.
\label{eq:bm}
\end{equation}
Here $V, V_i$ and $V_f$ are the four-velocities of the c.m., of initial
and of final spinor particle. The coefficient $\alpha$ and the normalization
coefficients $\sqrt{(M_i,M_i)}$ are equal to
\begin{eqnarray*}
\alpha = & -2\frac{(V,V_i+V_f)}{(V_i+V_f)^2} \;, \\
(M_1,M_1)=  &  2\left[1+(V_f,V_i)\right]\;,  \\
(M_2,M_2)=  &
2\frac{\left[1-(V,V_f)^2\right]\left[1-(V,V_i)^2\right]-
\left[(V_f,V_i)-(V,V_f)(V,V_i)\right]^2}{1+(V_f,V_i)}\;.
\end{eqnarray*}
In this case the corresponding cross section is 
equal to $|G^m|^2+|H^m|^2$. Thus there is the
unitary connection between $G^m,H^m$ and the canonical flip and non-flip 
c.m. $S$-matrix amplitudes
\[S=G^s+iH^s\hat{n} \;,\;\;\hat{n} \equiv n^k e_k\;,\;\;
\vec{n}=\frac{\vec{k}_i \times \vec{k}_f}{|\vec{k}_i \times
\vec{k}_f|}\;, \]
where $\vec{k}_{i,f}$ are initial and final c.m. momenta  \cite{Hohler1979}.
It is
\begin{eqnarray*}
G^m=\cos \frac{\varphi}{2}\;G^s -\sin \frac{\varphi}{2}\;H^s\;, \\
H^m=\sin \frac{\varphi}{2}\;G^s+\cos \frac{\varphi}{2}\;H^s\;,
\end{eqnarray*}
where the angle $\varphi$ is determined by
\begin{equation} e^{i\frac{\varphi}{2}}=
\sqrt{\frac{\omega_0 -\omega}{\omega_0 -\omega^{-1}}}\;,\;\;
\omega_0 \equiv \sqrt{\frac{(V,V_f)+1}{(V,V_f)-1}
\frac{(V,V_i)+1}{(V,V_i)-1}}\;,\;\;\;
\omega \equiv e^{i\theta }\;.
\label{eq:mu(w)}
\end{equation}
The $\theta$ is the c.m. scattering angle.

\section{Calculation of the transitive form factor F}

Let us chose the $p\Lambda$ c.m. system. Then 
\begin{eqnarray*}
p_b=(E_b,{\bf q}_b)\;,\; {\bf p}_ s={\bf \xi}\;,\;\;
(p_b-p_s)^2=2m^2-2E_bE_\xi+2q_b\xi x\;,\;\;     \\
E_b=\frac{s_{p\Lambda}+m^2-t}{2 W_{p\Lambda}}\;,\;\; 
q_b={\sqrt{[(W_{p\Lambda}+m)^2-t][(W_{p\Lambda}-m)^2-t]} \o 2W_{p\Lambda}}\;. 
\end{eqnarray*} 
Choosing z-axis along ${\bf q}_b$, taking into account 
$\d {\bf \xi}=\d \phi \d x \xi^2 \d \xi$ and integrating with
respect to azimuthal angle and $x$ we derive
\begin{eqnarray}
F(s_{p\Lambda},t) =\frac{g_{\pi}}{\sqrt{2}16\pi^2 q} 
\int_0^{\infty} \xi d\xi\;
\left[-\frac{1}{\xi^2-q_{cm}^2-i\epsilon}+\frac{1}{\xi^2+\beta^2}\right]\;.
\nonumber \\ 
\left[\ln \frac{2E_bE_\xi -2q_b\xi -2m^2+m_\pi^2}
{2E_bE_\xi +2q_b\xi -2m^2+m_\pi^2}-
\ln \frac{2E_bE_\xi -2q_b\xi -2m^2+\Lambda_\pi^2}
{2E_bE_\xi +2q_b\xi -2m^2+\Lambda_\pi^2}\right]\;,
\label{eq:3thFint}
\end{eqnarray}
where $q={W_{p\Lambda} \o m}q_b$. This integral can be computed if we replace the 
arguments of the  logarithms by the parabola with the roots coinciding
with the roots of these arguments. They are complex and equal to
\begin{eqnarray}
\xi(m_{\pi})_{\pm}=q_b(1-\frac{m_{\pi}^2}{2m^2})\pm 
iE_b\sqrt{1-(1-\frac{m_{\pi}^2}{2m^2})^2} \;, \nonumber  \\
\xi(\Lambda_{\pi})_{\pm}=q_b(1-\frac{\Lambda_{\pi}^2}{2m^2})\pm 
iE_b\sqrt{1-(1-\frac{\Lambda_{\pi}^2}{2m^2})^2} 
\end{eqnarray}
and we have to calculate
\bea
F(s_{p\Lambda},t) ={g_{\pi} \o 64\pi^2 q \sqrt{2}} 
\int_{-\infty}^{\infty} \d \xi\;
\left[-{1 \o \xi-q_{cm}-i\epsilon}-{1 \o \xi+q_{cm}+i\epsilon}
+{1 \o \xi+i\beta}+{1 \o \xi-i\beta}\right] \cdot \nn
\left[\ln {\xi-\xi(m_{\pi})_+ \o \xi+\xi(m_{\pi})_-}+
      \ln {\xi-\xi(m_{\pi})_- \o \xi+\xi(m_{\pi})_+}-
 \ln  {\xi-\xi(\Lambda_{\pi})_+ \o \xi+\xi(\Lambda_{\pi})_-}-
 \ln {\xi-\xi(\Lambda_{\pi})_- \o \xi+\xi(\Lambda_{\pi})_+}\right] \dqq
\l{eq:4thFint}
\eea
Here arguments of the each logarithms has singularities only
in the lower or upper half-plane. The first and the third logarithms are
free from singularities in lower half-plane and closing the contour for them 
in lower half-plane we meet the poles at $-q_{cm}-i\epsilon$ and $-i\beta$.
In contrast the second and forth logarithms pick up in upper half-plane
the poles at $q_{cm}+i\epsilon$ and $i\beta$. This yields
\begin{eqnarray}
F(s_{p\Lambda},t) =\frac{g_{\pi}}{\sqrt{2}16\pi q} i 
\nonumber \\ 
(-\ln \frac{-q_{cm}-i\epsilon-\xi(m_{\pi})_+}
{-q_{cm}-i\epsilon+\xi(m_{\pi})_-}
 +\ln \frac{-q_{cm}-i\epsilon-\xi(\Lambda_{\pi})_+}
{-q_{cm}-i\epsilon+\xi(\Lambda_{\pi})_-}+ \nonumber \\
      \ln \frac{-i\beta-\xi(m_{\pi})_+}{-i\beta+\xi(m_{\pi})_-}-
 \ln \frac{-i\beta-\xi(\Lambda_{\pi})_+}{-i\beta+\xi(\Lambda_{\pi})_-})=
\label{eq:5thFint}    \\
\frac{g_{\pi}}{16\pi q} i 
(-\ln \frac{-iq_{cm}-i\xi(m_{\pi})_+}{-iq_{cm}+i\xi(m_{\pi})_-}+
\ln \frac{-iq_{cm}-i\xi(\Lambda_{\pi})_+}{-iq_{cm}+i\xi(\Lambda_{\pi})_-}+
\nonumber \\
      \ln \frac{\beta-i\xi(m_{\pi})_+}{\beta+i\xi(m_{\pi})_-}-
 \ln \frac{\beta-i\xi(\Lambda_{\pi})_+}{\beta+i\xi(\Lambda_{\pi})_-})\;.
\nonumber
\end{eqnarray}
Defining
\begin{eqnarray}
\alpha_1=E_b\sqrt{1-(1-\frac{m_{\pi}^2}{2m^2})^2} \;,&\;\;
\alpha_2=E_b\sqrt{1-(1-\frac{\Lambda_{\pi}^2}{2m^2})^2}\;, \nonumber \\
\xi_{1\pm}=q_b(1-\frac{m_{\pi}^2}{2m^2})\pm q_{c.m.} \;,&\;\;
\xi_{2\pm}=q_b(1-\frac{\Lambda_{\pi}^2}{2m^2})\pm q_{c.m.} \;,  \\
C_1=1\;,&\;\;C_2=-1\;,  
\end{eqnarray}
we can write the final expression in the form similar to that of Laget
\begin{eqnarray}
F(s_{p\Lambda},t) =\frac{g_{\pi}}{32\pi q \sqrt{2}} \sum_j C_j \nonumber \\
\left[-i\ln \frac{\xi^2_{j+}+\alpha^2_j}{\xi^2_{j-}+\alpha^2_j}-
2\arctan \frac{\xi_{j+}}{\alpha_j}-2\arctan \frac{\xi_{j-}}{\alpha_j}+
4\arctan \frac{\xi_{j+}+\xi_{j-}}{2(\beta+\alpha_j)}\right]\;.
\end{eqnarray}

\end{document}